\documentclass[aps,showpacs,twocolumn]{revtex4-1}

\usepackage[english]{babel}
\usepackage[utf8]{inputenc}
\usepackage{amsmath}
\usepackage{amssymb}
\usepackage[caption=false]{subfig}
\usepackage{amssymb}
\usepackage{epsfig}
\usepackage{graphicx}
\usepackage{amsmath}
\usepackage{array,color}
\usepackage{natbib}

\usepackage[usenames,dvipsnames]{xcolor}
\definecolor{forestgreen}{rgb}{0.11,0.54,0.15}
\definecolor{purple}{rgb}{0.62,0.10,0.96}
\definecolor{dockerblue}{rgb}{0.11,0.56,0.98}
\definecolor{freeblue}{rgb}{0.25,0.41,0.88}

\usepackage[pdftex,plainpages=false,colorlinks=true,linkcolor=Red, citecolor=blue, urlcolor=blue]{hyperref}

\begin{document}

\title{Pseudogap, van Hove Singularity, Maximum in Entropy and Specific Heat for Hole-Doped Mott Insulators}
\author{A. Reymbaut$^{1,2}$, S. Bergeron$^{1,2}$, R. Garioud$^{1,2}$, M. Th\'{e}nault$^{2}$, M. Charlebois$^{1,2}$, P. S\'{e}mon$^3$ and A.-M. S. Tremblay$^{1,2,4}$*}
\affiliation{
$^1$Institut quantique, Universit\'{e} de Sherbrooke, Sherbrooke, Qu\'{e}bec, Canada J1K 2R1 \\
$^2$D\'{e}partement de physique and RQMP, Universit\'{e} de Sherbrooke, Sherbrooke, Qu\'{e}bec, Canada J1K 2R1 \\
$^3$Computational Science Initiative, Brookhaven National Laboratory, Upton, NY 11973-5000, USA \\
$^4$Canadian Institute for Advanced Research, Toronto, Ontario, Canada, M5G 1Z8
}
\date{\today}
\begin{abstract}
The first indication of a pseudogap in cuprates came from a sudden decrease of NMR Knight shift at a doping-dependent temperature $T^*(\delta)$. 
Since then, experiments have found phase transitions at a lower $T^*_\text{phase}(\delta)$.  
Using plaquette cellular dynamical mean-field for the square-lattice Hubbard model at high temperature, where the results are reliable, we show that $T^*(\delta)$ shares many features of $T^*_\text{phase}(\delta)$. 
The remarkable agreement with several experiments, including quantum critical behavior of the electronic specific heat, supports the view that the pseudogap is controlled by a finite-doping extension of the Mott transition.  
We propose further experimental tests.  
\end{abstract}
\email{Corresponding author andre-marie.tremblay@usherbrooke.ca}
\pacs{71.30.+h, 74.25.Dw, 71.10.Fd}
\maketitle



\section{Introduction}

Below a doping-dependent temperature $T^*$, early studies of cuprate high temperature superconductors found a decrease in NMR Knight shift \cite{Alloul:1989, Berthier:1996,Curro:1997,Timusk:1999,Kawasaki:2010}. 
This freezing of uniform spin fluctuations, a thermodynamic quantity, became the first signature of what is widely referred to as the pseudogap, one of the remaining challenges for theory.  
With time, another definition of the pseudogap became more popular. Polarized neutron diffraction \cite{Fauque:2006,Mook:2008,Mangin-Thro:2015}, Nernst effect measurements \cite{Daou:2010}, ultrasound measurements \cite{Shekhter:2013}, terahertz polarimetry \cite{Lubashevsky:2014} and optical anisotropy measurements \cite{Zhao:2017} report that the prototypical YBa$_2$Cu$_3$O$_y$ undergoes a thermodynamic phase transition that breaks time-reversal, spatial inversion, two-fold rotational, four-fold rotational and mirror symmetries below a doping-dependent temperature $T^*_\text{phase}$ that is distinctly lower than $T^*$ at low doping. This suggests that phase transitions are a consequence of the pseudogap first observed in NMR, not the cause~\cite{Yazdani:2010,Sordi:2012}.  

In this paper, we address the nature of the pseudogap that was first found in NMR.  
We focus mostly on thermodynamic signatures \textit{at high temperature}, where cluster generalizations of dynamical mean-field theory provide a reliable theoretical tool~\footnote{An 8 site DCA calculation of the Knight shift~\cite{Chen:2017,Wu:2018} for a smaller value of $U$ than that considered here, nevertheless gives a temperature dependence of the $T^*$ line quite comparable to ours.}.
The remarkable agreement that we find with several experiments supports the view that the high-temperature physics of the pseudogap is controlled by a finite-doping extension of the Mott transition that includes superexchange effects~\cite{Sordi:2010}. We propose further experimental tests to investigate that hypothesis. The connection between Mott transition and pseudogap was also suggested by calculations in smaller clusters~\cite{Stanescu_tJ:2004}.   

Although recent experimental results usually relate to the lower-temperature $T^*_\text{phase}(\delta)$ line, where $\delta$ denotes hole doping, the $T^*_\text{phase}(\delta)$ and $T^*(\delta)$ lines are almost parallel, which suggests that they are related.  Experimental results for $T^*_\text{phase}(\delta)$ can be classified into two families. 

The first family of results identifies the main features of the $T^*_\text{phase}$ line. 
Recent Hall measurements on YBa$_2$Cu$_3$O$_y$ (YBCO), La$_{2-x}$Sr$_x$CuO$_4$ (LSCO) and La$_{1.6-x}$Nd$_{0.4}$Sr$_x$CuO$_4$ (Nd-LSCO) highlighted a sharp jump in carrier density with increasing hole doping $\delta$ across a material-dependent critical doping $\delta^*_\text{phase}$ at which the $T^*_\text{phase}$ line suddenly drops \cite{Badoux:2016, Laliberte:2016, Collignon:2017}. 
This drop, also observed in Raman scattering experiments on Bi$_{2}$Sr$_{2}$CaCu$_2$O$_{8+\delta}$ (Bi2212) \cite{Loret:2017}, occurs between a low-doping antiferromagnetic Mott insulating regime \cite{Ando:2004} and a high-doping metallic regime \cite{Mackenzie:1996}. 
Moreover, the specific heat divided by temperature $C/T$ scales logarithmically as a function of temperature around this sharp transition in LSCO, Nd-LSCO and La$_{1.8-x}$Eu$_{0.2}$Sr$_x$CuO$_4$ (Eu-LSCO) \cite{Michon:2019}, which strengthens its interpretation as a finite temperature extension of an underlying quantum critical point. 

The second family of results regarding the $T^*_\text{phase}$ line identifies how its main features vary across different materials under external parameters, and how they relate to the rest of the phase diagram. 
The Nernst effect measurements of Ref.~\cite{Cyr_Choiniere:2018} establish that while both the $T^*_\text{phase}$ line's slope and the position of $\delta^*_\text{phase}$ differ between different parent compounds (YBCO and LSCO), only the position of $\delta^*_\text{phase}$ changes with chemical pressure within the same family of compounds (LSCO, Nd-LSCO and Eu-LSCO). 
Furthermore, it has been shown in Ref.~\cite{NDL:2018} that $\delta^*_\text{phase}$ shifts towards lower doping in Nd-LSCO under applied pressure. 
This unexpected shift seems to be driven by a corresponding shift in $\delta_\text{VH}$, the doping at which a van Hove singularity appears in the local density of states. 
Since this Lifshitz transition in cuprates translates into a change of the Fermi surface from hole-like to electron-like with increasing doping, as determined by angle-resolved photoemission spectroscopy (ARPES) \cite{Kondo:2004,Chang:2008,Yoshida:2009,Matt:2015}, this suggests that the pseudogap cannot open on an electron-like Fermi surface, so that the condition 
\begin{equation}
\delta^*_\text{phase} \leq \delta_\text{VH}
\label{Eq_conjecture}
\end{equation}
should always be satisfied. 
For instance, Ref.~\cite{Loret:2017} reports $\delta^*_\text{phase} \simeq \delta_\text{VH}$ in Bi2212. 
Apart from the van Hove singularity, experiments performed on calcium-doped YBCO and Bi2212 report a vertical regime of maximum electronic entropy in the phase diagram close to $\delta^*_\text{phase}$ \cite{Tallon:2001,Tallon:2004,Tallon:2007,Tallon_unpublished}. This observation is consistent with the behavior of thermopower when interpreted in terms of entropy~\cite{Particle-hole_Phillips:2010,Thermopower_Garg_Phillips_2011}.
Finally, the critical regime around $\delta^*_\text{phase}$ also corresponds to the zone in the phase diagram where superconductivity is most resilient to strong magnetic fields \cite{Grissonnanche:2014}, hinting that it may actually nurture superconductivity. 
This link between both phases of matter has been the subject of many past studies \cite{SordiSuperconductivityPseudogap:2012, Chen:2013, Gull:2013, Cyr_Choiniere:2018}.

While much is already known about the $T^*_\text{phase}$ line, much less is known regarding the fate of the NMR $T^*$ line with increasing doping, except for the fact that raw data for LSCO is consistent with  a sudden drop of $T^*$ at a doping $\delta^*$ close to $\delta^*_\text{phase}$ where $T^*_\text{phase}$ also drops~\cite{Nakano:1994}. 

However, most theoretical works study the $T^*$ line, focusing on the two-dimensional single-band Hubbard model on a square lattice \cite{Gutzwiller:1963,Hubbard:1963,Kanamori:1963}. 
In addition, cluster extensions of dynamical mean-field theory (DMFT), such as cellular dynamical mean-field theory (CDMFT) and the dynamical cluster approximation (DCA) \cite{Maier:2005,KotliarRMP:2006,LTP:2006} using various quantum impurity solvers \cite{Hirsch:1986,Rubtsov:2005, Gull:2011,werner:2006,WernerMillis:2006,haule:2007i,Gull:2007,Gull_AUX:2008}, 
have shown that the Hubbard model captures many features of the superconductivity and pseudogap of cuprate compounds \cite{Lichtenstein:2000,Jarrell:2001, Civelli:2005, LTP:2006, Macridin:2006, kyung:2006b, Haule:2007, Ferrero:2009, Gull:2009, Sakai:2009,Werner:2009,Gull:2010,Gull:2013,Gunnarsson:2015}. 
The same applies to its strong-interaction limit when correlated hopping is neglected: the $t$-$J$ model \cite{Haule:2002,Haule:2003,Bittner:2018}. 
DCA studies~\cite{Mikelsons:2009,Jarrell_Quantum_Critical:2009}  found a maximum in entropy and a critical logarithmic scaling for the specific heat $C(T)$ at a critical doping. 
Moreover, Refs.~\cite{Gunnarsson:2015, Maier:2016, Chen:2017} established that pseudogap formation and superconducting pairing are both linked through short-ranged spin physics \cite{Scalapino:1986, Maier:2008, Dahm:2009, Kyung:2009, Scalapino_RMP:2012, Nishiyama:2013, SenechalResilience:2013, Reymbaut:2016}. 
Closer to aforementioned experiments, Refs.~\cite{Wu:2018, Braganca:2018} found an unambiguous link between pseudogap formation and Fermi surface topology through the pole-like feature of the electronic self-energy, conjecturing that Eq.~\eqref{Eq_conjecture} for $\delta^*_\text{phase}$ should also hold for $\delta^*$.

Finally, a set of CDMFT studies spanning almost a decade has highlighted the existence of a finite-doping phase transition, here dubbed the ``Sordi transition''\cite{Sordi:2010, Sordi:2011}, that acts as an organizing principle for the $T$-$\delta$ phase diagram of cuprates \cite{Sordi:2012,Fratino:2016_2}. 
This transition is a finite-doping extension of the first-order Mott transition found at half-filling, separating pseudogap and correlated metallic phases in the normal-state phase diagram (where superconductivity is not allowed) \cite{Sordi:2010, Sordi:2011}. 
In this normal-state phase diagram, the Sordi transition ends at finite $T$ at a second-order critical point above which it extends as a Widom line \cite{XuStanleyWidom:2005} in the high-temperature crossover regime. 
Even if the low-temperature normal-state phase diagram is usually metastable, hidden by more ordered states, the associated high-temperature crossover regime can be observed in a doped organic compound thanks to magnetic frustration \cite{Oike:2014}. 
More specifically, in both doped Mott insulators \cite{Sordi:2012, Sordi:2013, Fratino:2016_2} and doped charge-transfer insulators \cite{Fratino:2016_1}, the Knight shift pseudogap $T^*$ line is parallel to the Widom line and appears as a high-temperature precursor of this crossover line. $T^*(\delta)$ drops to zero precipitously at $\delta^*$ in the vicinity of the Sordi transition. 
In addition, these CDMFT studies retrieve the aforementioned maximum in entropy close to $\delta^*$ \cite{Sordi:2011} and a van Hove-like singularity at higher doping \cite{Sordi:2012}. 


\section{Methods} 
\label{Sec_methods}

Here we study the two-dimensional Hubbard model on a square lattice
with $t$ the nearest-neighbour hopping, $t^\prime$ the next-nearest-neighbour hopping and $U$ the on-site Coulomb repulsion. The hopping $t=1$ serves as the energy unit, $k_B \equiv 1$ and $\hbar \equiv 1$. This model is implemented with CDMFT, where a 2$\times$2 cluster of sites~\cite{Kotliar:2001,Haule:2007} is dynamically coupled to a bath of non-interacting electrons by a frequency-dependent hybridization function. The quantum impurity problem is solved with continuous-time quantum Monte Carlo in the hybridization expansion \cite{werner:2006,WernerMillis:2006,haule:2007i,Gull:2007}. Note that close to half-filling, where finite-size effects are expected to be largest because of antiferromagnetism, systematic studies up to 5\% doping at $T=0.06\ t$ show that 2$\times$2 clusters give accurate results \cite{Sakai:2012}. All details regarding our various physical criteria can be found in the Supplemental Material~\footnote{In the Supplemental Material we explain how we find $T^*(\delta)$ and $\delta^*$ from the magnetic susceptibility, how we do analytic continuation to find the van Hove line $T_\text{VH}(\delta)$, how we use a Maxwell relation to find the location of the maximum of entropy and how we find the specific heat. We also explain how error bars are estimated.}.

\section{Results}
\label{Sec_results}
\subsection{NMR $T^*$, van Hove singularity and maximum of the entropy}

\begin{widetext}

\begin{figure}[h!]
\begin{center}
\includegraphics[width=0.48\textwidth]{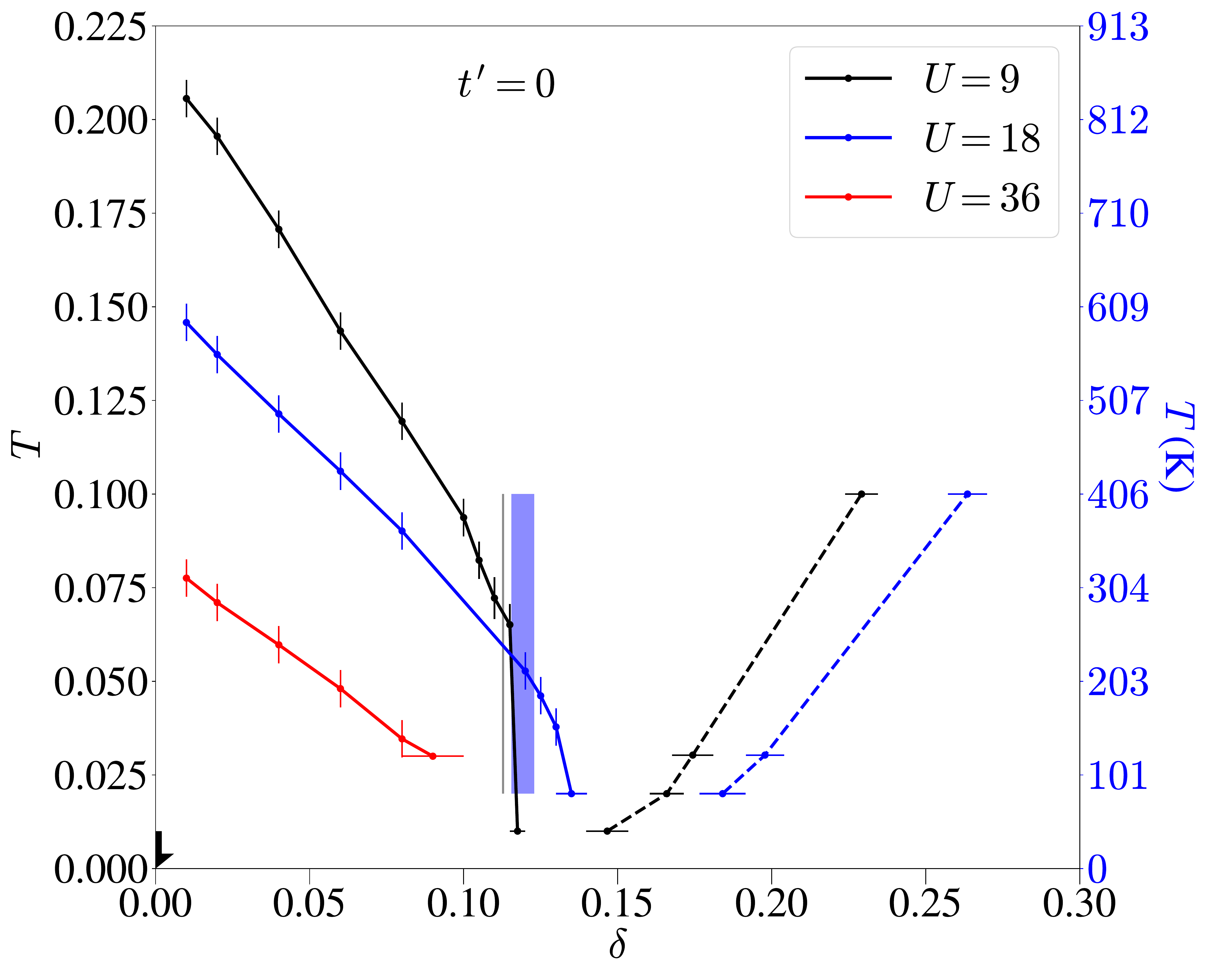}
\hspace*{\fill}
\includegraphics[width=0.48\textwidth]{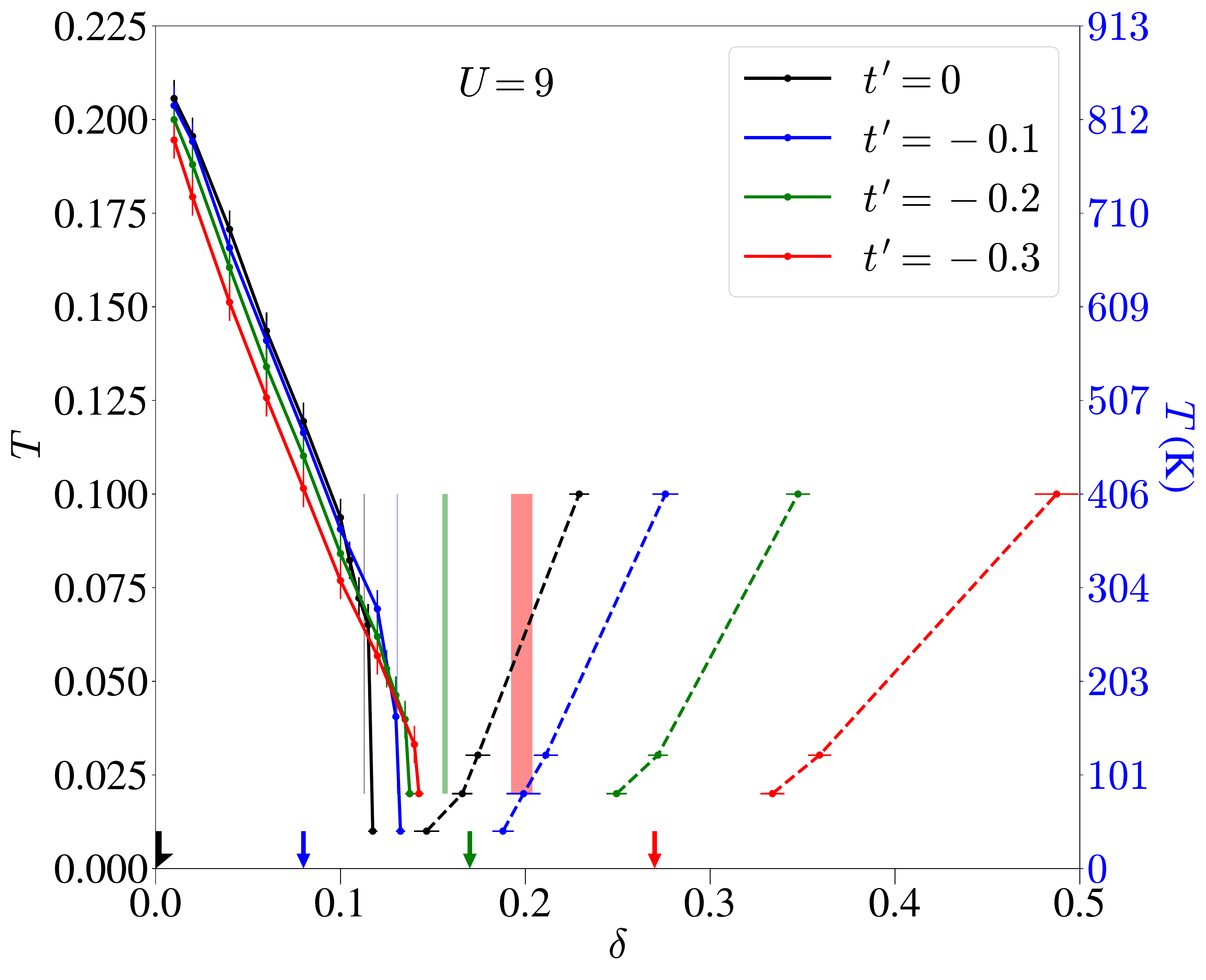}
\caption{For a given value of $U$ at $t^\prime=0$ (left) or finite $t^\prime$ at $U=9$ (right), the pseudogap $T^*$ line is represented by a solid line, the van Hove line by a dashed line, and the zone of maximum entropy by a filled rectangle. The arrows indicate the positions of the van Hove singularities in the non-interacting case for different $t'$. Severe sign problems sometimes occurred at low temperature or large doping, especially for $U=36$. The blue temperature axis was obtained from $t\simeq 350$ meV.}
\label{Fig_Phase_Diagram}
\end{center}
\end{figure}

\end{widetext}

Fig.~\ref{Fig_Phase_Diagram} (left) shows the normal-state phase diagram at $t^\prime = 0$ for $U=9$, $18$ and $36$. Note that our values of $T^*$ for a 2$\times$2 cluster match the ones found in larger cluster studies with comparable $U$ \cite{Maier_Scalapino:2018}.
The case $U=36$ suffered from severe sign problems but is still useful to capture the evolution of $T^*(\delta\to 0)$ as a function of $U$.
Within the Hubbard model, pseudogap formation arises from singlet formation due to superexchange, which we retrieve in this figure~\cite{Haule:2007,Sordi:2012}.
Indeed, these correlations are well described by the $t$-$J$ limit of the Hubbard model at strong interaction and low doping, with superexchange $J = 4t^2/U$. 
Going from $U=9$ to $U=18$, $T^*(\delta\to 0)$ does not actually decrease by a factor of $2$, because the interaction strength is not large enough for the $t$-$J$ limit to be valid. 
However, $T^*(\delta\to 0)$ does decrease by a factor of $2$ between $U=18$ and $U=36$. 
In other words, $T^*(\delta\to 0)$ scales with $J$ \cite{Haule:2002,Haule:2003,Stanescu_tJ:2004} for large values of $U$, confirming short-range correlations due to superexchange as the origin of the pseudogap. 
Furthermore, the order of magnitude of $T^*(\delta\to 0)$ at $U=9$ agrees with the experimental values reported in Refs.~\cite{Johnston:1989,Nakano:1994,Curro:1997} for YBCO and LSCO. 
If we assume that the value of the N\'eel temperature is a measure of $J$, this is consistent with the experimental slope of the $T^*_\text{phase}$ lines of YBCO and LSCO cited in Ref.~\cite{Cyr_Choiniere:2018}. Putting $U=36$ aside in Fig.~\ref{Fig_Phase_Diagram} (left), since sign problems appear at low temperature, we find at intermediate doping that the $T^*$ line drops suddenly at a critical doping $\delta^*$: 
the smaller the value of $U$, the larger the drop of $T^*$. The increase of $\delta^*$ with $U$ follows the trend found for the location of the Sordi transition \cite{Fratino:2016_2}.

At finite temperature in the presence of interactions, Fig.~\ref{Fig_Phase_Diagram} also shows dashed lines in the $T$-$\delta$ plane that indicate the maximum of the single-particle local density of states at the Fermi level. This is not the usual non-interacting van Hove singularity, but it is adiabatically connected to it. 
We see on the figure that the dashed lines (the van Hove lines $T_\text{VH}(\delta)$) are pushed to higher doping compared to the location of the van Hove singularity in the non-interacting case (arrows). 
Consequently, the experimental conjecture Eq.~\eqref{Eq_conjecture} for $\delta^*_\text{phase}$ seems, in our calculations, to extend to $\delta^*$ since $\delta^* \leq \delta_\text{VH}$.

The location of the zone of maximum of entropy in the $T$-$\delta$ plane, represented by a colored rectangle, is also pushed by interactions to higher doping. 
The displacement of the zone of maximum entropy towards higher doping $\delta$ with increasing $U$ would be consistent with its coincidence with the end of the coexistence region of the Sordi transition \cite{Sordi:2011}, whose $\delta$ also increases with $U$. This coindence has however been proven only for $U=6.2$. 

It seems paradoxical that with increasing $U$ the $T^*$ line extends to higher doping but appears at lower temperatures. 
This comes from antagonistic effects of $U$ on singlet formation. 
On the one hand, local magnetic moments increase with $U$, leading to better-defined singlets at low temperature. 
On the other hand, the decrease of $J$ with increasing $U$ makes these singlets less resilient to thermal fluctuations. 

Fig.~\ref{Fig_Phase_Diagram} (right) shows the normal-state phase diagram at $U=9$ for $t^\prime = 0$, $-0.1$, $-0.2$ and $-0.3$. The dependence on $t^\prime$ of $\delta^*$ follows the trend found for the location of the Sordi transition (see Supplemental Material of Ref.~\cite{Fratino:2016_2}). 
Remarkably, two main experimental observations regarding $\delta^*_\text{phase}$ and $T^*_\text{phase}(\delta)$ are also found for $\delta^*$ and $T^*(\delta)$ in this figure. 
Indeed, one experiment~\cite{Cyr_Choiniere:2018} finds that changing the value of $t^\prime$ through chemical pressure does not affect the initial slope of the $T^*_\text{phase}$ line at low doping but $\delta^*_\text{phase}$ monotonically moves towards higher doping with increasing values of $\vert t^\prime \vert$. 
Moreover, another experiment~\cite{NDL:2018} shows that applied pressure pushes $\delta^*_\text{phase}$ to lower doping while calculations in the same paper~\cite{NDL:2018} find a concomitant decrease of $\vert t^\prime\vert$. 

Interactions push the van Hove line $T_\text{VH}(\delta)$ to higher doping than the non-interacting van Hove singularity,  as in the $t'=0$ case. The effect of $t'$ at fixed $U$ agrees with Ref.~\cite{Wu:2018}: 
the larger $\vert t^\prime \vert$, the less $U$ pushes the van Hove line away from the non interacting case (colored arrows in Fig.~\ref{Fig_Phase_Diagram}). 
This is consistent with the decreasing influence of interactions at large doping. 
Besides, the experimental conjecture Eq.~\eqref{Eq_conjecture} for $\delta^*_\text{phase}$ extends once again to $\delta^*$ as $\delta^* \leq \delta_\text{VH}$. 
Finally, the zone of maximum entropy also moves towards higher doping with increasing $\vert t^\prime \vert$. 
More importantly, it is farther and farther away from both the Widom and the van Hove lines, as suggested in yet unpublished work \cite{Sordi_unpublished}. \\

\subsection{Critical scaling of the specific heat}
\begin{widetext}

\begin{figure}[h!]
\begin{center}
\includegraphics[width=0.48\textwidth]{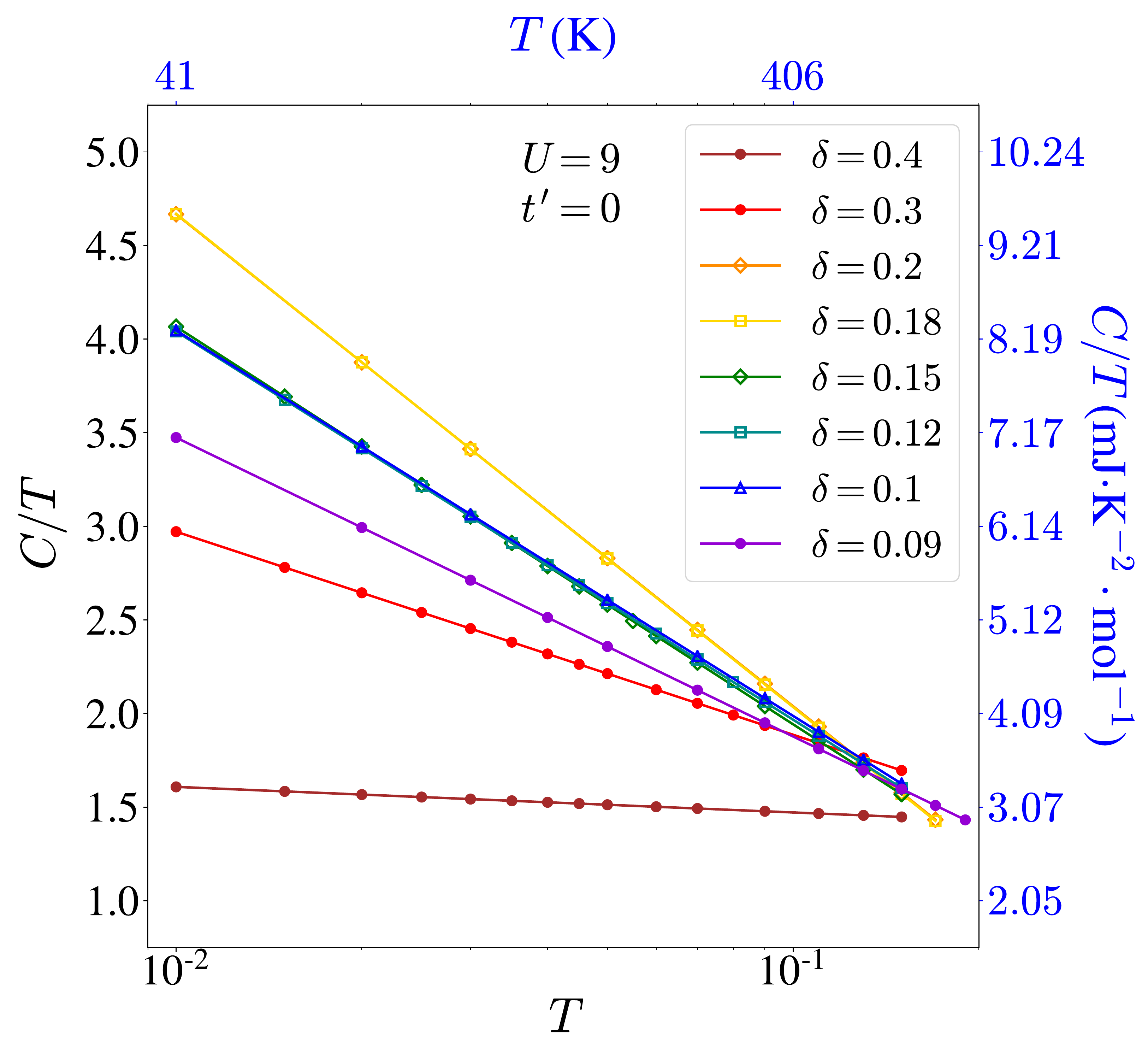}
\hspace*{\fill}
\includegraphics[width=0.48\textwidth]{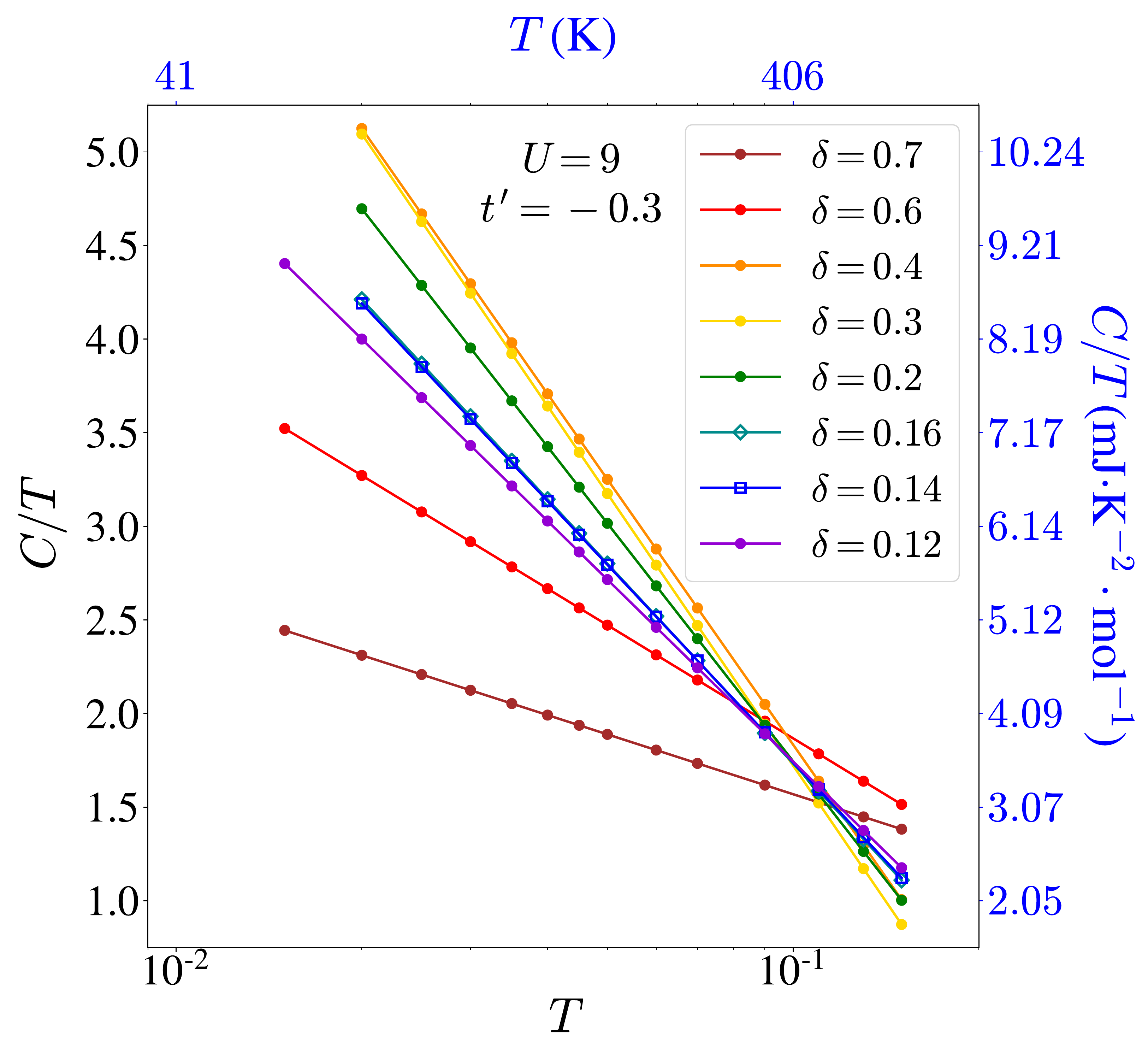}
\caption{Specific heat divided by temperature as a function of temperature for different dopings at $U=9$, $t^\prime=0$ (left) and $U=9$, $t^\prime=-0.3$ (right). After extracting the total energy of the system and fitting it as $E_\text{tot} = a + bT^2\ln T + cT^2$, the specific heat is extracted as its derivative with respect to temperature, yielding $C/T = 2b\ln T + (b+2c)$. The solid lines are guides to the eye. The temperature in Kelvin and specific heat axes (in blue) were generated using $t\simeq 350$ meV.}
\label{Fig_Spec_heat}
\end{center}
\end{figure}

\end{widetext}

Let us finally discuss critical scaling of the specific heat. 
We computed the total energy~\cite{Fratino:2016_2} and fitted it with $E_\text{tot} = a + bT^2\ln T + cT^2$. 
The specific heat is the derivative of the fitted total energy with respect to temperature, yielding $C/T = 2b\ln T + (b+2c)$, presented on a semi-log plot in Fig.~\ref{Fig_Spec_heat} for $U=9$, $t^\prime=0$ (left) and for $U=9$, $t^\prime=-0.3$ (right). 
Since the goodness of the energy fit holds over a wide range of dopings, as shown in the Supplemental Material~\cite{Note1}, so does the critical scaling of $C/T$. 
A typical cuprate hopping amplitude of $t\simeq 350$ meV converts $C/T$ to a value that is of the same order as that found experimentally~\cite{Michon:2019}. 
However, our lowest temperature $T=0.01\,t$ is about $40$ K. 
This minimal temperature is much higher than the experimental range $0.5$ K to $10$ K in Ref.~\cite{Michon:2019} where a critical fan with a sharp peak in the doping dependence of $C/T$ at fixed $T$ appears. 
Instead, we find that at higher temperature the value of $C/T$ is essentially the same for a wide range of parameters around the critical point $\delta^*\sim 0.12$ for $t'=0$ and $\delta^*\sim 0.15$ for $t'=-0.3$ that we determined from $T^*$. 

This observation can be understood by taking experimental studies in heavy fermions~\cite{Lohneysen:1996,Lohneysen:2007} as typical examples of what is expected for the quantum critical behavior of $C/T$. 
Heavy fermions exhibit an extremely low effective Fermi energy so that the high-temperature limit of the electronic specific heat is accessible without phonon contamination.
It is found that at high temperature $C/T$ is indeed logarithmic and very weakly dependent on parameters. 
The equivalence established in Ref.~\cite{NDL:2018} between applying pressure in experiments and decreasing $\vert t^\prime\vert$ in calculations enables us to establish another analogy with Refs.~\cite{Lohneysen:1996,Lohneysen:2007}: 
at a given doping, the slope of $C/T$ as a function of temperature on the semi-log plot becomes flatter with decreasing $\vert t^\prime\vert$, like we find by comparing the two plots in Fig.~\ref{Fig_Spec_heat}.
\footnote{Logarithmic temperature dependence of $C/T$ is also found in the two-impurity Kondo problem \protect\cite{Affleck_Ludwig_1992} 
where long-wavelength spin fluctuations are unimportant. The singlets that are found in the pseudogap phase of plaquette CDMFT calculations \cite{Sordi:2012,Haule:2007} may lead to similar behavior.} 
When doping increases beyond the critical doping, $C/T$ again becomes insensitive to doping for a small doping range, reflecting the presence of the van Hove line $T_\text{VH}(\delta)$.
Upon increasing doping sufficiently, $C/T$ gradually becomes temperature independent, as expected in a Fermi liquid. 

\section{Conclusion}
\label{Sec_conclusion}

In summary, consistent with most earlier theoretical studies, we find that the NMR pseudogap $T^*$ line~\footnote{In Bi2201, the NMR $T^*$~\cite{Kawasaki:2010} seems to coincide with $T^*_\text{phase}$ as determined from ARPES and resistivity.} shares many features of the experimental $T^*_\text{phase}$ line~\cite{Loret:2017,Michon:2019,Cyr_Choiniere:2018,NDL:2018,Tallon:2001,Tallon:2004,Tallon:2007,Tallon_unpublished}: 
a) $T^*$ drops precipitously to zero at a doping $\delta^*$; 
b) $T^*$ near half-filling scales like superexchange $J=4t^2/U$ at large $U$; 
c) Changes in the band structure modify primarly the value of $\delta^*$ and it is possible to see this effect experimentally by applying pressure; 
d) The doping where the van Hove line extrapolates at low temperature, $\delta_\text{VH}(T\to 0)$, seems to satisfy the inequality Eq.~(\ref{Eq_conjecture}) in the $T=0$ limit;
e) The high-temperature specific heat is consistent with expectations for high-temperature quantum critical behavior; 
f) The doping where entropy is maximum is almost independent of $T$.


Our contribution for theory is that quantum critical behavior of the electronic specific heat is connected to doped Mott-insulating behavior. 
Indeed, the extrapolated position of the quantum critical point seems to have the same dependence on $t'$ as the Sordi transition, which is continuously connected to the Mott transition at half-filling.
The critical-point temperature for the Sordi transition is known to decrease rapidly with increasing $U$ and should be investigated further. 

Additional experiments on the NMR $T^*$ are called for to verify our observation that the $T^*$ line is a necessary condition for the appearance of the experimental $T^*_\text{phase}$ line. That $T^*_\text{phase}$ line occurs at lower temperature and is thus more sensitive to details of the Hamiltonian. 
Another prediction for experiment is that the doping where entropy is maximum depends strongly on band parameters, an effect that could be measured with combined specific-heat and pressure experiments. 


\paragraph*{Acknowledgments} 
We are grateful to S. Verret and M.-H. Julien for useful discussions, to O. Simard for help, and to Louis Taillefer for comments on the manuscript. The work of P.S. was supported by the U.S. Department of Energy, Office of Science, Basic Energy Sciences as a part of the Computational Materials Science Program. This research was undertaken thanks in part to funding from the Canada First Research Excellence Fund. 
This work has been supported by the Natural Sciences and Engineering Research Council of Canada (NSERC) under grant RGPIN-2014-04584, and by the Research Chair in the Theory of Quantum Materials. Simulations were performed on computers provided by the Canadian Foundation for Innovation, the Minist\`ere de l'\'Education des Loisirs et du Sport (Qu\'ebec), Calcul Qu\'ebec, and Compute Canada.

%

\end{document}


\title{Pseudogap, van Hove Singularity, Maximum in Entropy and Specific Heat for Hole-Doped Mott Insulators}
\author{A. Reymbaut$^{1,2}$, S. Bergeron$^{1,2}$, R. Garioud$^{1,2}$, M. Th\'{e}nault$^{1,2}$, M. Charlebois$^{1,2}$, P. S\'{e}mon$^2$ and A.-M. S. Tremblay$^{1,2,3}$}
\affiliation{
$^1$Institut quantique, Universit\'{e} de Sherbrooke, Sherbrooke, Qu\'{e}bec, Canada J1K 2R1 \\
$^2$D\'{e}partement de physique and RQMP, Universit\'{e} de Sherbrooke, Sherbrooke, Qu\'{e}bec, Canada J1K 2R1 \\
$^3$Canadian Institute for Advanced Research, Toronto, Ontario, Canada, M5G 1Z8
}
\maketitle
\begin{widetext}
In this Supplemental Material we explain how we find $T^*(\delta)$ and $\delta^*$ from the magnetic susceptibility, how we do analytic continuation to find the van Hove line $T_\text{VH}(\delta)$, how we use a Maxwell relation to find the location of the maximum of entropy and how we find the specific heat and how it varies as a function of doping. We also explain how error bars are estimated.\\

\paragraph*{Pseudogap temperature $T^*$}

For given values of $U/t \equiv U$ and $t^\prime/t \equiv t^\prime$, we define the pseudogap temperature $T^*$ in a way similar to Refs.~\cite{Alloul:1989,Sordi:2012}, \textit{i.e.} as the temperature at which the spline interpolation of the uniform spin susceptibility at zero frequency $\chi^{(\omega=0)}$ is maximized along a temperature sweep at fixed doping:
\begin{equation}
T^*(\delta) = \underset{T}{\mathrm{argmax}}[\chi^{(\omega=0)}(T,\delta)]\, .
\label{Criterion_1}
\end{equation}
The associated error bar is given by the temperature step between two consecutive points in the sweep. Criterion~\eqref{Criterion_1} is illustrated in Fig.~\ref{Fig_PG}. 

\begin{figure}[hb!]
    \begin{center}
    \includegraphics[width=0.49\textwidth]{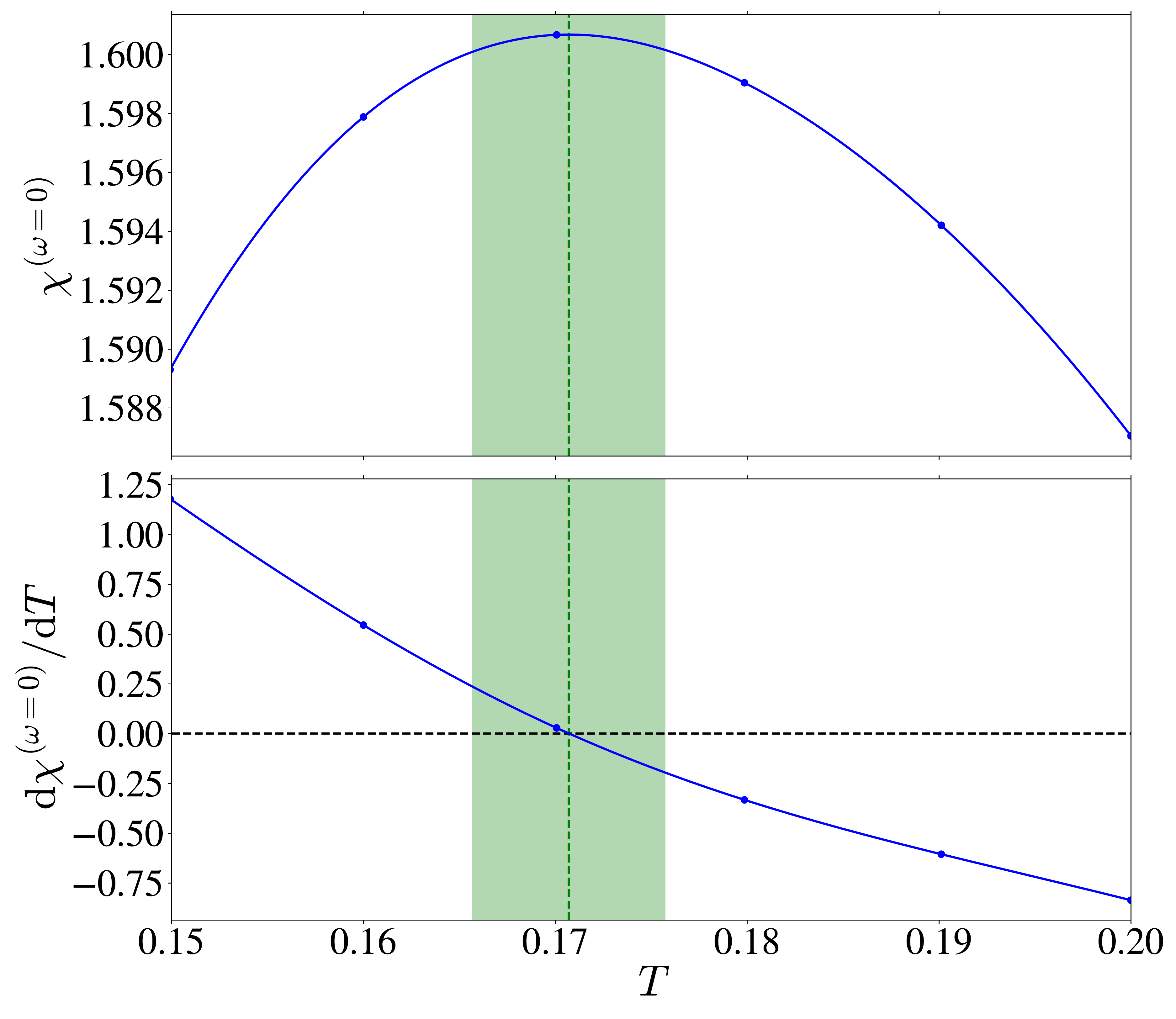}
    \caption{Top: Spline interpolation of the data points giving the uniform spin susceptibility at zero frequency $\chi^{(\omega=0)}$ as a function of temperature at fixed doping. The temperature where the spline interpolation is maximum coincides with the Knight shift pseudogap temperature $T^*$ (vertical dashed line). The associated error bar is given by the temperature step between two consecutive points in the sweep (green span). Bottom: Derivative of the above spline interpolation with respect to temperature. The horizontal dashed line at zero serves as a guide to the eye to verify that the maximum in $\chi^{(\omega=0)}$ corresponds to a zero in its derivative.}
    \label{Fig_PG}
    \end{center}
    \end{figure}

\paragraph*{Critical doping $\delta^*$}

The critical doping $\delta^*$ is found by taking the average of the last doping at which $T^*$ can be found and the first doping at which it cannot be found. Its associated error bar joins these two dopings. This critical doping is represented in the phase diagram by a point $(\delta^*,T_\text{min})$, where $T_\text{min}$ is the minimum temperature that was investigated to find $T^*$ above $\delta^*$. \\

\paragraph*{Van Hove doping $\delta_\text{VH}$}

For given values of $U$ and $t^\prime$, we define the van Hove doping $\delta_\text{VH}$ as the doping at which the local density of states at zero frequency $\mathcal{A}_\text{loc}^{(\omega=0)}$ is maximized along a doping sweep at fixed temperature:
\begin{equation}
\delta_\text{VH}(T) = \underset{\delta}{\mathrm{argmax}}[\mathcal{A}_\text{loc}^{(\omega=0)}(T,\delta)]\, .
\label{Criterion_2}
\end{equation}
$\mathcal{A}_\text{loc}^{(\omega=0)}$ is obtained from polynomial extrapolation of the low Matsubara frequencies in the local Green's function. Criterion~\eqref{Criterion_2} is illustrated in Fig.~\ref{Fig_VH}. Higher-order extrapolations generate unphysical oscillations in the spline interpolation for $\mathcal{A}_\text{loc}^{(\omega=0)}$ and are not considered here. There is a direct equivalence between $\delta_\text{VH}(T)$ and $T_\text{VH}(\delta)$.

\begin{figure}[hb!]
    \begin{center}
    \includegraphics[width=0.49\textwidth]{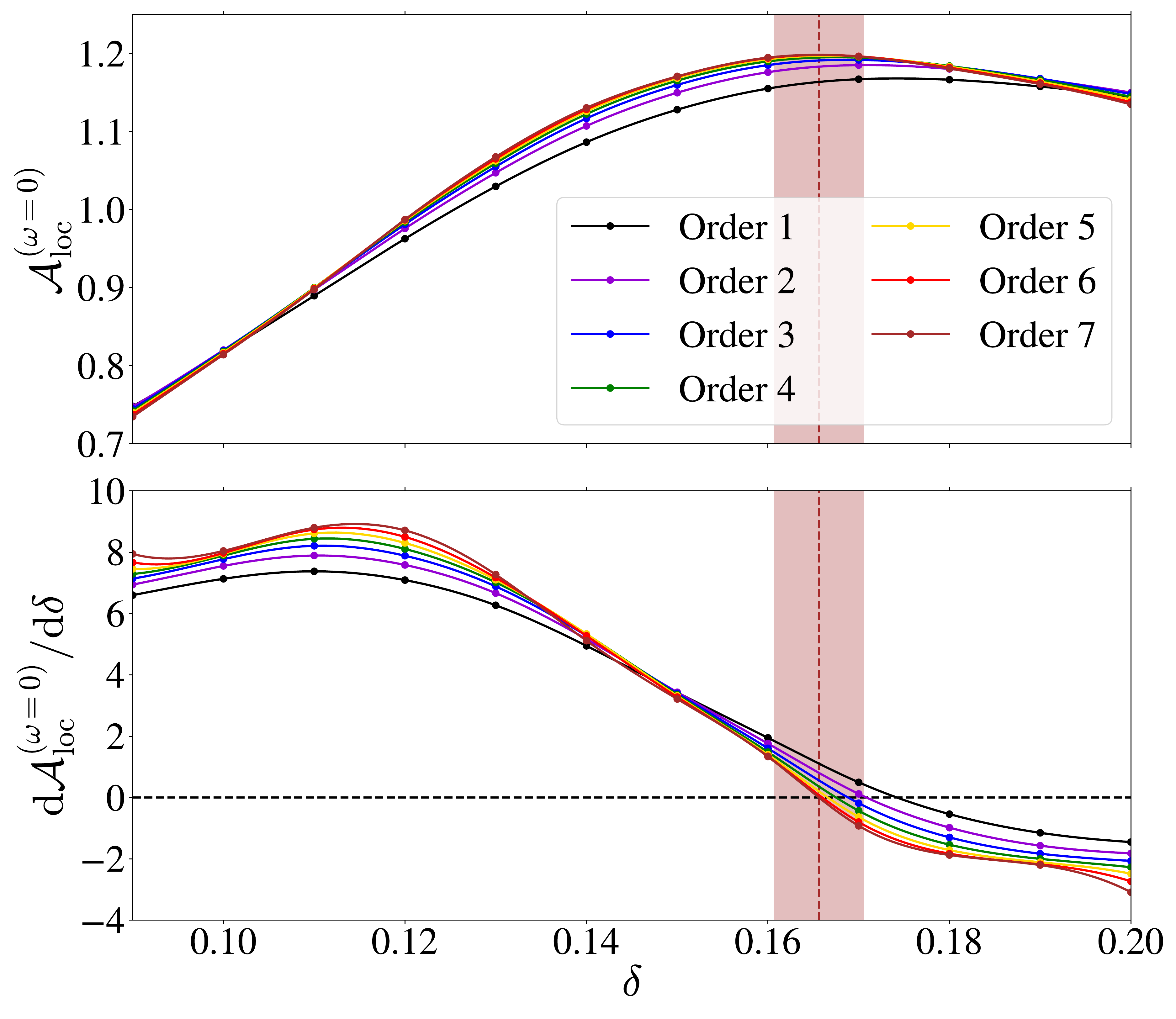}
    \caption{Top: Spline interpolation of the data points giving the uniform local density of states at zero frequency $\mathcal{A}_\text{loc}^{(\omega=0)}$ as a function of doping at fixed temperature for different orders of polynomial extrapolation of the low Matsubara frequencies in the local Green's function. The doping where the highest-order extrapolation is maximum defines the van Hove doping $\delta_\text{VH}$. The associated error bar is given by the doping step between two consecutive points in the sweep (colored spans). Bottom: Derivative of the above spline interpolation with respect to doping.}
    \label{Fig_VH}
    \end{center}
    \end{figure}
    
\newpage
\paragraph*{Zone of maximum entropy}

The zone of maximum in entropy is found using the same Maxwell relation as Ref.~\cite{Sordi:2011}:
\begin{equation}
\left. \frac{\partial s}{\partial \mu} \right\vert_T = \left. \frac{\partial n}{\partial T} \right\vert_\mu = 0\, ,
\end{equation}
where $s$ denotes the entropy per site. This criterion is illustrated in Fig.~\ref{Fig_maxS}.

\begin{figure}[h!]
    \begin{center}
    \includegraphics[width=0.49\textwidth]{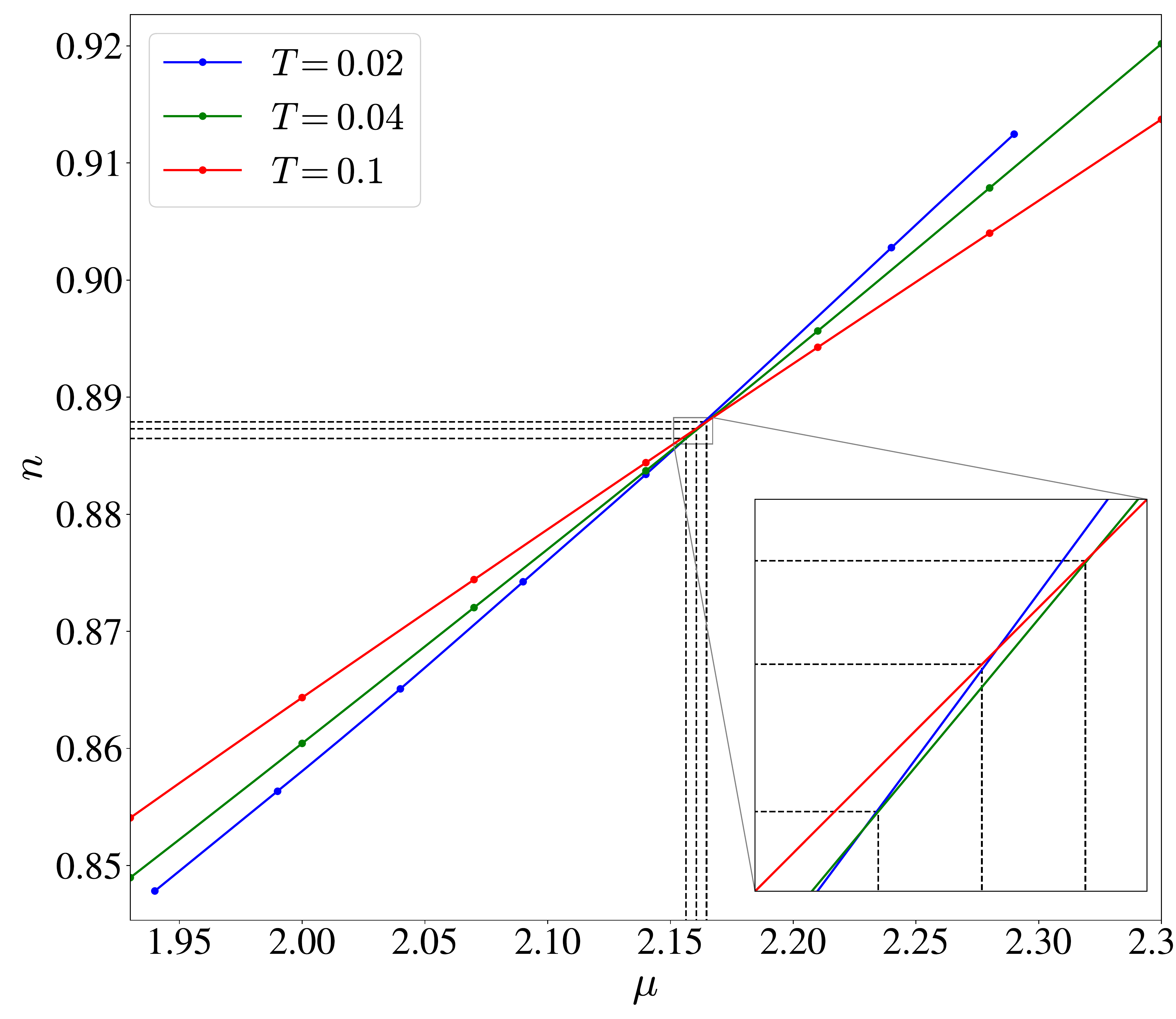}
    \caption{Occupancy $n$ as a function of chemical potentiel $\mu$ for different temperatures. To obtain a general zone of maximum entropy, we plot $n$ as a function of $\mu$ for $T=0.02$, $T=0.04$ and $T=0.1$ and isolate the three intersection points $(\mu_1,n_1)$, $(\mu_2,n_2)$ and $(\mu_3,n_3)$ between pairs of curves. The zone of maximum entropy is then represented in the phase diagram by the vertical rectangle whose opposite corners have coordinates $(\delta = 1-\mathrm{max}[n_1,n_2,n_3],T=0.02)$ and $(\delta=1-\mathrm{min}[n_1,n_2,n_3],T=0.1)$, thereby giving the most conservative estimate of this zone.}
    \label{Fig_maxS}
    \end{center}
    \end{figure}

\newpage
\paragraph*{Goodness of the total energy fit}

In order to extract the specific heat, we computed the total energy as Ref.~\cite{Fratino:2016_2} and fitted it with $E_\text{tot} = a + bT^2\ln T + cT^2$. The quality of this fit at various dopings is presented in Fig.~\ref{Fig_E}.


\begin{figure}[h!]
\begin{center}
\includegraphics[width=0.49\textwidth]{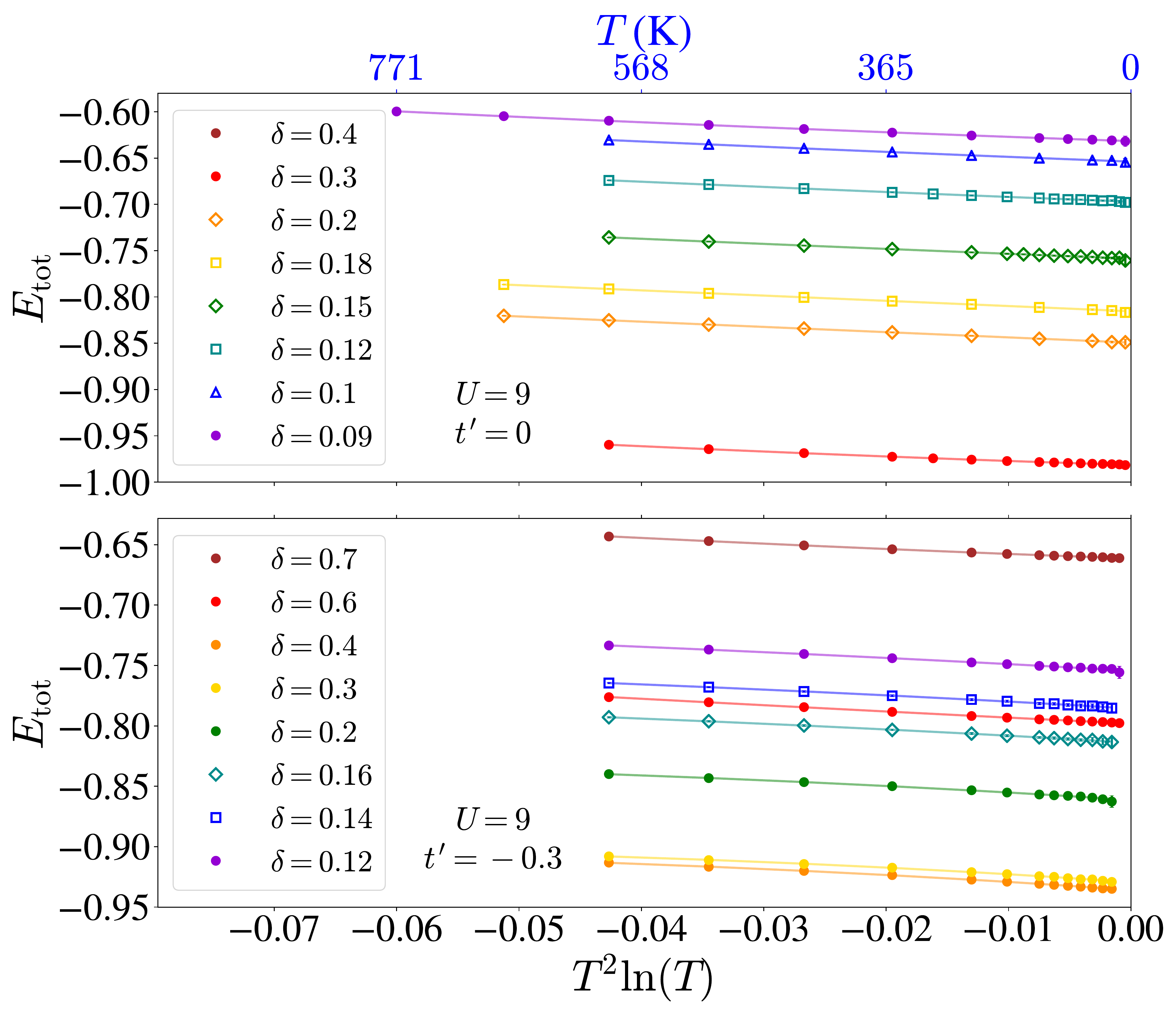}
\caption{Total energy as a function of $T^2\ln(T)$ (points) fitted as $E_\text{tot} = a + bT^2\ln T + cT^2$ (solid lines) for $U=9$, $t^\prime=0$ (top), and $U=9$, $t^\prime = -0.3$ (bottom). The real temperature axis (in blue) was generated using $t\simeq 350$ meV. Error bars, indicated on the figure, are usually smaller than the symbols. 
}
\label{Fig_E}
\end{center}
\end{figure}

\vspace{12pt}

\paragraph*{$C/T$ at low temperature as a function of doping}

Figure~\ref{Fig_C_T} shows the specific heat at low temperature as a function of doping. The maximum is more clearly visible in recent studies~\cite{Sordi:2019} done for values of $U$ closer to the critical $U$ for the Mott transition. In that case, the first-order transition and associated critical point can be clearly seen. The physics of the maximum becomes visible at higher temperature.

\begin{figure}[h!]
    \begin{center}
    \includegraphics[width=0.45\textwidth]{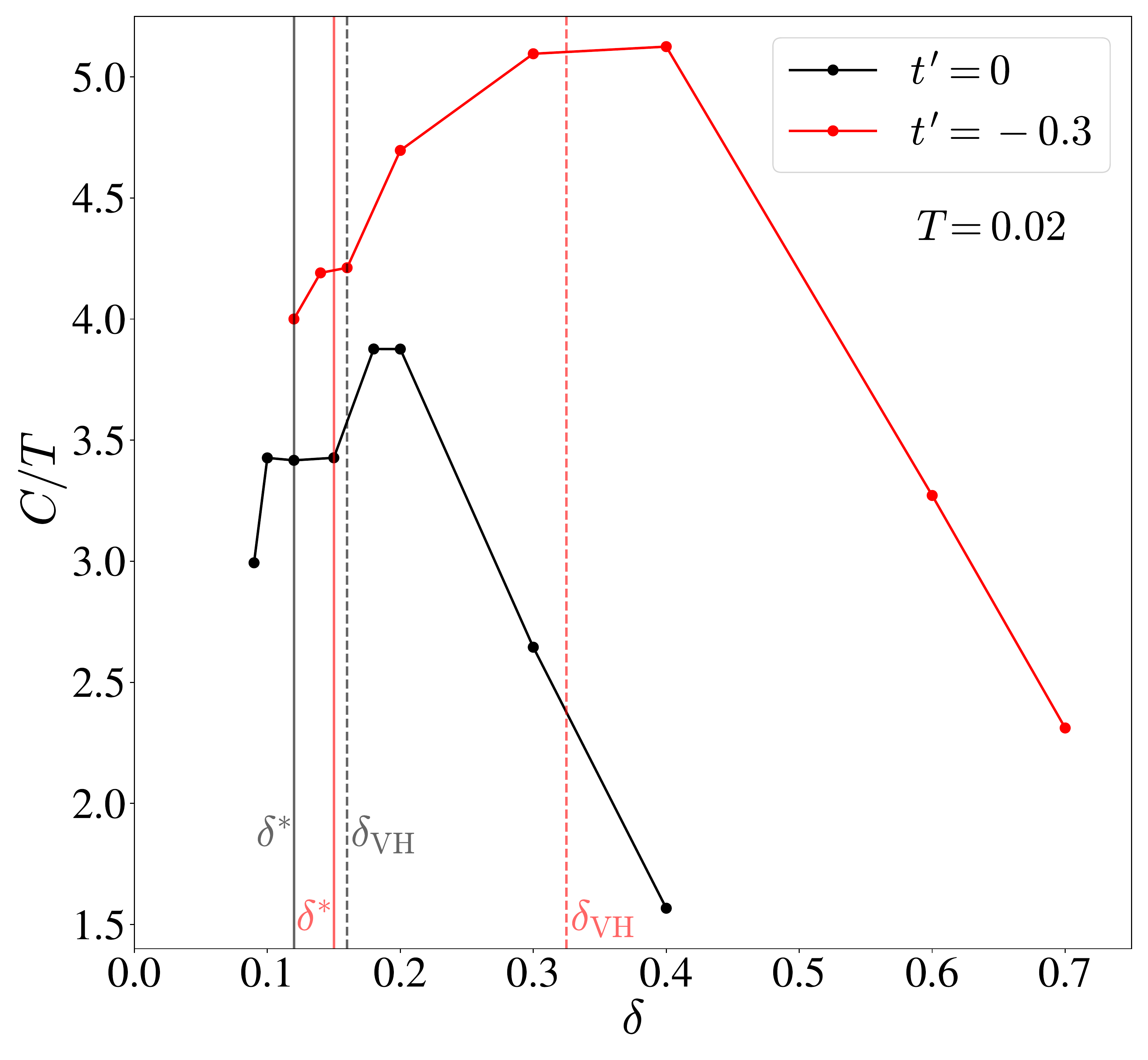}
    \caption{Value of $C/T$ as a function of doping at the lowest common temperature $T=0.02$ across the $t'=0$ and $t'=-0.3$ datasets at $U=9$. There is a local maximum near the critical doping $\delta^*$ and near the interaction-modified van Hove singularity. 
    }
    \label{Fig_C_T}
    \end{center}
    \end{figure}

\end{widetext}



%